\documentclass[letterpaper]{article} 
\usepackage{aaai2026}  
\usepackage{times}  
\usepackage{helvet}  
\usepackage{courier}  
\usepackage[hyphens]{url}  
\usepackage{graphicx} 
\usepackage{natbib}  
\usepackage{caption} 
\usepackage{algorithm}
\usepackage{algorithmic}

\usepackage{amsmath}
\usepackage{makecell}
\usepackage{amssymb}
\usepackage{xcolor}
\usepackage{pifont}
\definecolor{tred}{RGB}{255,0,0}
\definecolor{tgreen}{RGB}{0,128,0}
\usepackage{subfigure}
\usepackage{threeparttable}

\definecolor{myblue}{HTML}{1d71fe}
\definecolor{myred}{HTML}{c00000}
\definecolor{improvered}{RGB}{34, 139, 34}
\definecolor{decreaseblue}{RGB}{70, 130, 180}

\usepackage{booktabs} 
\usepackage{multirow} 

\usepackage{array}

\usepackage{newfloat}
\usepackage{listings}
\DeclareCaptionStyle{ruled}{labelfont=normalfont,labelsep=colon,strut=off} 
\floatstyle{ruled}
\newfloat{listing}{tb}{lst}{}
\floatname{listing}{Listing}
\title{Sentient: Detecting APTs Via Capturing Indirect Dependencies and Behavioral Logic}
\author {
    Wenhao Yan\textsuperscript{\rm 1,\rm 2},
    Ning An\textsuperscript{\rm 1,\rm 2},
    Wei Qiao\textsuperscript{\rm 1,\rm 2},
    Weiheng Wu\textsuperscript{\rm 1,\rm 2},
    Zhigang Lu\textsuperscript{\rm 1,\rm 2},
    Bo Jiang\textsuperscript{\rm 1,\rm 2},
    Baoxu Liu\textsuperscript{\rm 1,\rm 2},
    Junrong Liu\textsuperscript{\rm 1,\rm 2}\thanks{Corresponding author.}
}
\affiliations {
    \textsuperscript{\rm 1}Institute of Information Engineering, Chinese Academy of Sciences, Beijing, China\\
    \textsuperscript{\rm 2} School of Cyber Security, University of Chinese Academy of Sciences, Beijing, China\\
    \{yanwenhao,anning,qiaowei,wuweiheng,luzhigang,jiangbo,liubaoxu,liujunrong\}@iie.ac.cn
}

\usepackage{bibentry}

\begin{document}

\maketitle

\begin{abstract}
Advanced Persistent Threats (APTs) are difficult to detect due to their complexity and stealthiness. To mitigate such attacks, many approaches model entities and their relationship using provenance graphs to detect the stealthy and persistent characteristics of APTs. However, existing detection methods suffer from the flaws of missing indirect dependencies, noisy complex scenarios, and missing behavioral logical associations, which make it difficult to detect complex scenarios and effectively identify stealthy threats.
In this paper, we propose Sentient, an APT detection method that combines pre-training and intent analysis. It employs a graph transformer to learn structural and semantic information from provenance graphs to avoid missing indirect dependencies. We mitigate scenario noise by combining global and local information. Additionally, we design an Intent Analysis Module (IAM) to associate logical relationships between behaviors. Sentient is trained solely on easily obtainable benign data to detect malicious behaviors that deviate from benign behavioral patterns.
We evaluated Sentient on three widely-used datasets covering real-world attacks and simulated attacks. Notably, compared to six state-of-the-art methods, Sentient achieved an average reduction of 44\% in false positive rate(FPR) for detection.
\end{abstract}


\section{Introduction}
Advanced Persistent Threats are notorious for their stealth and complexity. Attackers meticulously plan their strategies to gain long-term control over the target system and remain dormant, causing significant damage to modern enterprises and institutions~\cite{alshamrani2019aptsurvey}. For example, the SolarWinds~\cite{solarwinds2020} attack in 2020, which compromised multiple U.S. government agencies and major private companies, exposed sensitive information and disrupted operations on a global scale. Thus, effective detection of APTs has been a focus of both industry and academia.

The detection of APTs remains a paramount concern in cybersecurity research and practice. Leveraging audit logs to mine attack traces is currently considered the most effective approach. However, traditional Intrusion Detection Systems (IDS) struggle to cope with complex and persistent attacks, resulting in insufficient detection accuracy~\cite{inam2023sok}. Recent studies have analyzed valuable information in provenance graphs, effectively utilizing the rich contextual data within audit logs to provide robust support for threat detection and attribution.

Existing provenance graph-based APT detection approaches can be categorized into three classes: \textbf{Statistical-based approaches}~\cite{wang2020path_de,liu2018towards,hassan2019nodoze} quantify anomalies by analyzing the rarity of events in provenance graphs. However, these approaches only focus on direct event connections while ignoring the deep semantics and hidden relationships in provenance graphs. \textbf{Rule-based approaches}~\cite{xiong2020conan,hassan2020tactical,milajerdi2019poirot,milajerdi2019holmes} rely on expert-defined rules to detect threats. However, these approaches require extensive prior domain knowledge that must be distilled by experts. \textbf{Learning-based approaches}~\cite{han2020unicorn,jia2024magic,yang2023prographer,rehman2024flash,wu2025brewing} employ machine learning techniques to identify anomalous behaviors and attack patterns in provenance graphs. However, attack events are often submerged in a large number of benign events, and complex dependency acquisition is limited by the receptive field of Graph Neural Networks (GNNs).

\begin{figure*}[ht]
    \centering
    \includegraphics[width=\linewidth]{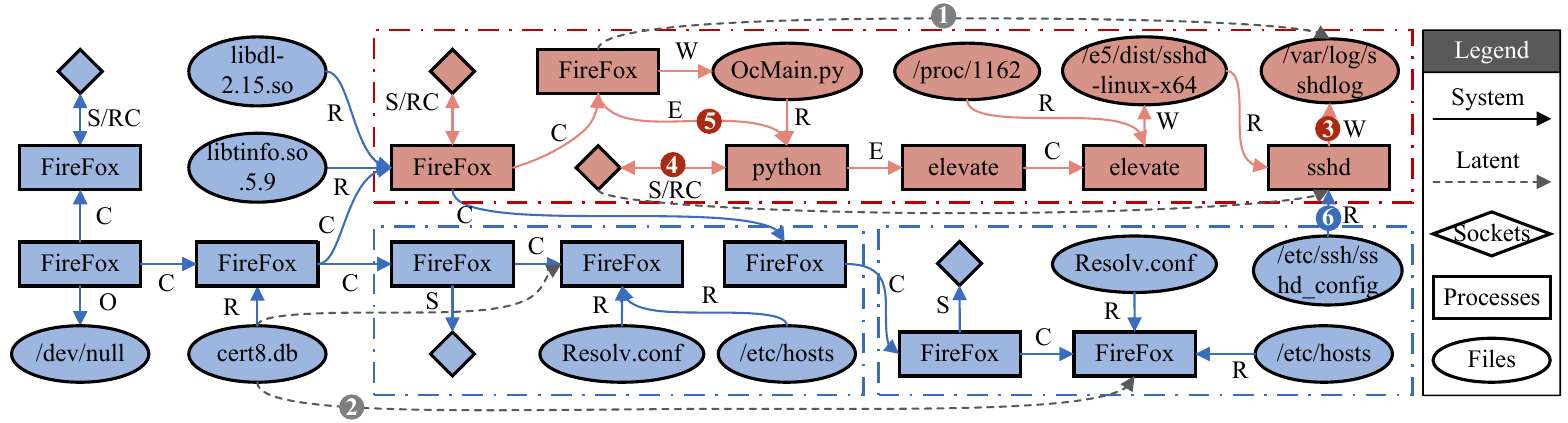}
    \caption{An example of an attack provenance graph from the DARPA E3 dataset. Red and blue denote malicious and benign, respectively. The red subgraphs highlight the core attack behavior, while the blue subgraphs represent benign DNS resolution activity. The system call type: R=Read, W=Write, O=Open, C=Clone, E=Execute, S=Send, and Rc=Receive.}
    \label{fig:Motivation}
\end{figure*}

Although previous research has achieved certain effectiveness in attack detection, current methods still face several challenges in complex scenarios. We illustrate these challenges through a real attack scenario, as shown in Figure~\ref{fig:Motivation}, where \ding{182} to \ding{187} are all labeled in the figure.

\ding{172} \textbf{Missing indirect dependencies.} Two entities connected by direct interaction edges have an associated relationship, but this does not mean that non-directly connected nodes lack associations. For example, the indirect dependencies shown in \ding{182} and \ding{183} are often overlooked by models.

\ding{173} \textbf{Noisy complex scenarios.} Infected entities continue to perform a large number of legitimate tasks (i.e., benign activities). Additionally, attackers may hide their attack behaviors by embedding benign substructures into their attack strategies. This causes detection models to be interfered with by weakly correlated activities, thereby producing erroneous results. For example, neighbor aggregation erroneously aggregates behaviors \ding{184} and \ding{187} that lack strong correlation, merely because they are spatially adjacent.

\ding{174} \textbf{Missing behavioral logical association.} Isolated system behaviors exhibit contextual diversity. For example, when considered independently, the sshd process writing to sshdlog represented by \ding{185} might appear as routine logging activity under normal circumstances, while the socket interaction of the Python process represented by \ding{186} might be interpreted as standard network communication behavior. However, the logical association of behaviors \ding{185}, \ding{186}, and \ding{187} makes the malicious intent of the sshd and Python processes evident. However, due to the large distance between \ding{184} and \ding{185}, \ding{186}, and the influence of GNNs receptive fields, existing methods struggle to associate them.

To address the aforementioned gaps, we propose Sentient, an APT detection methodology that integrates global and local information. In contrast to existing approaches, Sentient learns established benign behavioral patterns to identify anomalous behaviors that deviate from these benign patterns, without requiring prior knowledge of attack patterns. The pre-training module leverages a graph transformer incorporating semantic and positional encodings to attend to all pertinent node information within the provenance graph, rather than merely neighboring information, thereby enabling Sentient to capture indirect dependencies. The Intent Analysis module constructs denoised scenarios for each node to mitigate information interference and captures logical associations between events through enhanced Mamba2. Additionally, Sentient simplifies the defense process by clustering behaviors with similar intent.

We implemented Sentient and compared it against six existing state-of-the-art methods on three APT attack datasets. Sentient outperformed virtually all other methods, achieving 96\% accuracy and 99\% recall, while reducing the average false positive rate (FPR) by 44\%. We also validated Sentient's robustness against adversarial attacks.

In summary, this paper makes the following contributions.
\begin{itemize}
\item We propose Sentient, an accurate APT detection method for complex scenarios and stealthy attacks. It uses only easily obtainable benign data for training to learn normal behavioral intents and detect anomalous behaviors deviating from benign patterns.
\item We construct a pre-training module to mine fundamental structural and semantic information in graphs, avoiding the loss of indirect dependencies.
\item We combine global information to construct denoised scenarios for each node to avoid interference from noisy information.
\item We design an Intent Analysis Module (IAM) to learn logical associations between behaviors.
\item We conduct comprehensive evaluations using real-world datasets, and the results demonstrate the effectiveness and robustness of Sentient in detecting APTs.
\end{itemize}

\section{Related Work}
\subsection{APT Detection}
Based on the usage of audit log, current detection methods for APTs can be classified into three categories. We summarize their specific limitations in Table~\ref{tab:Obstacles}.
\begin{table}[tph]
    \centering
    \small
    \setlength{\tabcolsep}{1.2mm}
    \begin{tabular}{lcccc}
        \hline
        \textbf{Model} & \makecell{\textbf{Super-}\\ \textbf{vision}} & \textbf{\begin{tabular}[c]{@{}c@{}}
        \ding{172} Indirect\\Dependency\end{tabular}} & \textbf{\begin{tabular}[c]{@{}c@{}}\ding{173} Neighbor\\Denoising\end{tabular}} & \textbf{\begin{tabular}[c]{@{}c@{}}\ding{174} Logical\\Association\end{tabular}} \\
        \hline
        \textbf{Sentient} & B & \textcolor{tgreen}{\ding{51}} & \textcolor{tgreen}{\ding{51}} & \textcolor{tgreen}{\ding{51}} \\
        Unicorn & B & \textcolor{tred}{\ding{55}} & \textcolor{tred}{\ding{55}} & \textcolor{tred}{\ding{55}} \\
        Atlas & B, A & \textcolor{tred}{\ding{55}} & \textcolor{tred}{\ding{55}} & \textcolor{tgreen}{\ding{51}} \\
        ProGrapher & B & \textcolor{tred}{\ding{55}} & \textcolor{tred}{\ding{55}} & \textcolor{tred}{\ding{55}} \\
        FLASH & B & \textcolor{tred}{\ding{55}} & \textcolor{tred}{\ding{55}} & \textcolor{tred}{\ding{55}} \\
        Slot & B, A & \textcolor{tgreen}{\ding{51}} & \textcolor{tgreen}{\ding{51}} & \textcolor{tred}{\ding{55}} \\
        \hline
    \end{tabular}
    \caption{Comparison of APT detection approaches. Within column supervision, B indicates benign data, A refers to attack data.}
    \label{tab:Obstacles}
\end{table}

\textbf{Statistical-based methods}, such as ~\cite{liu2018towards,hassan2019nodoze,wang2020path_de} analyzes the distributional characteristics of system interaction behaviors and quantifies suspiciousness based on interaction frequency between system entities. These methods assume that attacks are strongly correlated with uncommon behaviors. However, rare events are not always anomalous, and this approach cannot capture deep semantic and event correlations, leading to inaccurate detection.

\textbf{Rule-based methods}, like ~\cite{hossain2017sleuth,milajerdi2019holmes} construct attack rule databases from previous attacks and detect threats by matching audit logs against these rules. ~\cite{xiong2020conan,hossain2020combating} define rules using expert knowledge for anomaly scoring, generating alerts when scores exceed thresholds. However, these methods are limited by threat intelligence and cannot detect unknown threats, while struggling to adapt dynamically to changing network environments.

\textbf{Learning-based methods} model system behavioral patterns in provenance graphs to evaluate system anomalies. For example, Unicorn~\cite{han2020unicorn} employs graph similarity matching to detect abnormal graphs. Atlas~\cite{alsaheel2021atlas} applies lemmatization and word embeddings to generate sequences and uses LSTM networks to predict whether sequences are related to attacks. ProGrapher~\cite{yang2023prographer} uses graph embedding to construct subgraph representations and detects threats based on logical relationships between subgraphs. Threatrace~\cite{wang2022threatrace} and Flash~\cite{rehman2024flash} reconstruct node types to determine anomalies. Slot~\cite{qiao2025slot} integrates graph reinforcement learning, dynamically aggregating neighbor information to learn correlated behaviors for threat detection.
While node-level detection effectively identifies anomalies, current methods aggregate neighbor information to learn behavioral associations, which ignores indirect dependencies between non-adjacent nodes while introducing ambiguous dependencies between unrelated system behaviors. 

\subsection{Graph Transformer}
The Transformer with attention mechanisms has achieved significant success in various fields, including Natural Language Processing~\cite{vaswani2017attention} and Computer Vision. In Graph Transformers~\cite{chennagphormer}, global attention is typically computed, allowing each node to attend to all other nodes, regardless of edge connections. This enables Graph Transformers to effectively capture indirect dependencies.

\subsection{State Space Models}
General state space models (SSMs), such as Hidden Markov Models and RNNs, process sequences by recurrently updating hidden states and generate outputs by combining hidden states with inputs. Structured State Space Models (S4) improve computational efficiency through reparameterization.
Building upon S4, Mamba~\cite{gu2023mamba} and Mamba2~\cite{dao2024transformers} introduce a data-dependent selection mechanism to eliminate gradient vanishing or explosion problems that occur during long sequence processing.

\section{Threat Model}
Building upon prior research in APT detection ~\cite{wang2022threatrace,cheng2024kairos,rehman2024flash,jia2024magic}, our study focuses on scenarios involving stealthy and complex attack activities. During the initial intrusion phase, attackers attempt to blend malicious activities with legitimate background data to obscure their attack intentions.
Similar to prior works, this model assumes the auditing system provides accurate, integrity-protected activity records. By ensuring log integrity with tamper-resistant storage, these logs serve as a reliable foundation for behavioral analysis and threat detection.

\begin{figure*}[ht]
    \centering
    \includegraphics[width=\linewidth]{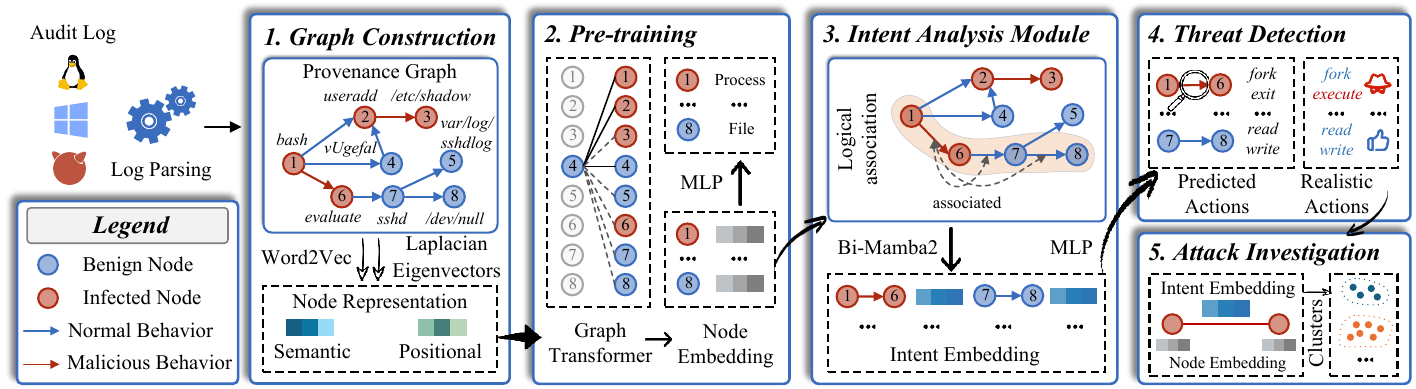}
    \caption{Overview of Sentient's Architecture.}
    \label{fig:Overview}
\end{figure*}
\section{Methodology}
As shown in Figure~\ref{fig:Overview}, Sentient comprises five key components: 
\noindent \textbf{(1) Graph Construction.} Sentient constructs provenance graphs from system logs and builds initial node representations based on semantic information and graph topological structure. 
\noindent \textbf{(2) Pre-training.} Sentient reconstructs node key information to comprehensively understand semantic and structural information in provenance graphs, generating ideal node embeddings that capture complex graph information. 
\noindent \textbf{(3) Intent Analysis Module.} Sentient combines node embeddings to construct behavioral sequences, providing a denoised environment for event association. It employs a bidirectional selection mechanism to identify logical associations between behaviors and embedded representations of behavioral intent are constructed.
\noindent \textbf{(4) Threat Detection.} Anomalies are detected by verifying the similarity between reconstructed and actual edge information. 
\noindent \textbf{(5) Attack Investigation.} Behaviors with similar intents are clustered and attack stories are reconstructed.
\subsection{Graph Construction}
Sentient constructs provenance graphs from audit data collected by logging infrastructures such as Windows ETW, Linux Audit, and CamFlow~\cite{camflow}. In the graph $G = (V, E)$, nodes $v \in V$ represent entities (e.g., files, processes) and edges $e \in E$ denote system interactions (e.g., write, open).

To better represent interaction scenarios and leverage rich attributes between system logs and entities, Sentient employs Word2Vec~\cite{mikolov2013efficient} to project attributes (e.g., process name, file path) into low-dimensional vector representations while preserving correlations between node attributes. This semantic encoding ensures that files with similar functionalities (e.g., \texttt{/bin/vim}, \texttt{/bin/nano}) are projected into closer vector spaces, while distinct files (e.g., \texttt{/bin/vim} and \texttt{/bin/python3}) are projected into more discrete spaces.

Previous studies \cite{ying2021transformers,huang2024good} have shown that the Laplacian eigenvectors of a graph possess strong positional awareness capabilities. We apply Laplacian transformation to the adjacency matrix of the provenance graph to construct Laplacian eigenvectors, which are then used as positional representations for the nodes. The eigenvectors are derived from the graph Laplacian matrix factorization:
\begin{align}
L &= I - D^{-\frac{1}{2}} A D^{-\frac{1}{2}}, \\
L v_i &= \beta_i v_i, \quad i = 1, 2, \dots, k+1.
\end{align}
where \( A \) is the adjacency matrix and \( D \) is the degree matrix. The \( k \) smallest non-trivial eigenvectors, denoted as \( v_i \) for each node \( i \), are used as positional encodings for the nodes, with \( \beta_i \) representing the eigenvalue associated with \( v_i \).

\subsection{Pre-training}
A complete user behaviors or attack activities typically involve multiple system calls, which manifest as multi-hop node relationships in provenance graphs. However, existing learning-based detection methods predominantly employ GNNs for neighbor message aggregation, which causes indirect relationships between system behaviors to be overlooked. In contrast to these approaches, Sentient incorporates attention mechanisms to attend to all node information, thereby capturing indirect behavior (as shown \ding{182} and \ding{183} in Figure~\ref{fig:Motivation}). 

Specifically, Sentient combines semantic encoding \(\alpha\) and Laplacian positional encoding \(\beta\) to form initial embeddings. The objective is to create a representation that incorporates both semantic information and the graph's topological structure, calculated as follows:
\begin{align}
h_i^0 &= \sigma((A^0\alpha+a^0) + (B^0\beta+b^0))\;.
\end{align}
where \(A^0\), \(a^0\), \(B^0\) and \(b^0\) are parameters of the linear layers, and the \(h_i^0\) is the initial representation of the node \(i\). We use attention to mine indirect behaviors between nodes to capture complicated relationships in complete events. The node update equations for layer \(\ell\) are as follows:
\begin{align}
\hat{h}_i^{\ell+1} &= O_h^{\ell} \underset{k=1}{\overset{H}{\Big\Vert}} \left( \sum_{j \in \mathcal{N}_i} w_{ij}^{k,\ell} V^{k,\ell} h_j^{\ell} \right), \\
w_{ij}^{k,\ell} &= \text{softmax}_j \left( \frac{Q^{k,\ell} h_i^{\ell} \cdot K^{k,\ell} h_j^{\ell}}{\sqrt{d_k}} \right),
\end{align}
where $Q^{k,\ell}$, $K^{k,\ell}$, $V^{k,\ell}$ are the query, key, and value matrices for the $k$-th attention head at layer $\ell$, $H$ is the number of attention heads, and $\Vert$ denotes concatenation.
The final node representation is obtained through residual connections and normalization:
\begin{align}
h_i^{\ell+1} &= \mathrm{Norm}\left( \mathrm{Norm}(h_i^\ell + \hat{h}_i^{\ell+1}) + \mathrm{FFN}(\hat{h}_i^{\ell+1}) \right),
\end{align}
where $\mathrm{Norm}$ denotes layer normalization and $\mathrm{FFN}$ is a feed-forward network.

Finally, the node embedding $h_i$ captures indirect relationships between system events, which facilitates the comprehension of complete user activities.
To learn user behaviors from large-scale log data and guide the generation of node embeddings that capture indirect relationships between system events, we construct a node key information reconstruction task based on node types (e.g., files, processes, sockets). Sentient employs a multi-layer perceptron (MLP) to generate probability distributions over node types:  $f(h_i) = \text{MLP}(h_i).$
Where $h_i$ represents the embedding of node $i$ generated by Equation (7), and $f(h_i)$ denotes the probability vector for node type classification. However, node type distributions in system logs often exhibit significant imbalance. To counteract this, we employ a weighted cross-entropy loss function.
\begin{equation}
\begin{split}
Loss(y, f(x)) = &-\sum_{i=1}^n \big(w_i y_i \log(f(x)_i)\\
& + (1 - y_i) \log(1 - f(x)_i) \big).
\end{split}
\end{equation}
where $y$ represents the true node type vector, $n$ denotes the number of classes, $w_i$ is the weight for the $i$-th class, and $\log$ denotes the natural logarithm.

\subsection{Intent Analysis Module}

As mentioned earlier, neighbor noise interferes with the accurate capture of correlated behaviors. To address this challenge, Sentient employs a pre-trained Graph Transformer model to generate node embeddings $h$ for provenance graphs, and utilizes random walk algorithms~\cite{nikolentzos2020random} to construct behavior sequences $\lambda$ on the provenance graph. The $i$-th behavior sequence is defined as $\lambda_i=\{e_1,e_2,\ldots,e_W\}$, where $W$ represents the maximum sequence length. Each behavior $e_t$ in the sequence is represented as $[h_{\phi(e_t)};h_{\psi(e_t)}]$.
where $\phi(e)$ denotes the function to obtain the source node of behavior $e$, and $\psi(e)$ denotes the function to obtain the destination node of behavior $e$.

Mining logical associations between behaviors is essential for understanding behavioral intent. Sentient builds a Bidirectional Mamba2 (Bi-Mamba2) architecture to enhance bidirectional logical association capabilities on behavioral sequences. Mamba2 has shown significant effectiveness in NLP applications, particularly sentiment analysis \cite{qu2024survey}, due to its ability to capture complex sequential dependencies.

Specifically, Sentient takes the behavioral sequence representation as the initial input $\lambda^{(0)}$ for Bi-Mamba2. Each layer of Bi-Mamba2 processes information in both forward and backward directions simultaneously. For the $\ell$-th layer, the computation is formulated as:
\begin{align}
\lambda^{\ell+1} &= \mathbf{F}(\mathbf{E}(\lambda^{\ell}) + \mathcal{R}(\mathbf{E}(\mathcal{R}(\lambda^{\ell}))), \lambda^{\ell}).
\end{align}
where $\mathbf{E}(\lambda^{\ell})$ is $\{y_1, \ldots, y_L\}$, $\mathcal{R}$ denotes sequence reversal operation, and $\mathbf{F}$ represents a combination function that incorporates residual connections and feed-forward neural network layers. The variable $y$ at each layer is computed as:
\begin{align}
    s_t = A s_{t-1} + B x_t, \quad y_t = C s_t + D x_t,
\end{align}
where $x_t$ denotes the input at the current time step $t$, $s_{t-1}$ represents the hidden state at the previous time step. The $s_t$ represents the derivative of the hidden state, i.e. how the state is evolving. The matrices $A$, $B$, $C$, and $D$ govern state transition, input-to-state mapping, state-to-output projection, and direct feedthrough, respectively.

Sentient captures logical relationships between behaviors through the IAM and generates behavioral intent embeddings $h_{e_t}$ for different scenarios. 

\subsection{Threat Detection}
Subsequently, Sentient employs an MLP to reconstruct the action distribution $a_t$ for different behaviors: $\mathbf{P}(a_t) = \text{MLP}(h_{e_t})$.

\noindent \textbf{Learning Benign Patterns.} Sentient mask critical behavioral information (i.e., read, write, execute operations), the model learns normal user behavioral patterns using only \textbf{benign} log. Sentient minimizes the reconstruction error (RE) between the predicted probability vector $P(a_t)$ and the observed action $L(a_t)$ from benign provenance graphs:
\begin{align}\label{eq:re}
\text{RE} = \text{CrossEntropy}(\mathbf{P}(a_t), L(a_t)).
\end{align}
where $L(a_t)$ is a one-hot vector with probability 1 for the actual edge type of $e_t$ and 0 elsewhere.

\noindent \textbf{Anomaly Detection.} In the detection phase, Sentient applies its learned understanding of benign behavioral patterns from the training phase. It uses a Graph Transformer to analyze system and lateral movement behaviors. Subsequently, IAM captures logical relationships, generating embedded representations of behaviors across various scenarios. An MLP then reconstructs these behavioral actions.

To quantify similarity, the reconstructed actions are compared to the observable actual actions by computing the RE using Equation \ref{eq:re}. Behaviors with an RE exceeding a predefined threshold are classified as malicious, while those below the threshold are categorized as normal. The predefined threshold is based on the average RE value learned during the benign behavior learning phase. We found that using the average RE value plus 1.5 times the standard deviation yields superior detection performance.

\subsection{Attack Investigation}
Despite certain detection capabilities, threat detection methods inevitably produce false positives and false negatives, making security analysis heavily dependent on manual review~\cite{false_postivte}. The vast volume of audit logs makes it challenging for analysts to process them effectively.

Sentient constructs node embeddings capable of capturing complex scenario information through a pre-trained graph model. The IAM captures behavioral intent embeddings $h_e$ for behavior $e$. Sentient concatenates these with node embeddings $h_{\phi(e)}$ and $h_{\psi(e)}$, representing the source and destination nodes of behavior $e$ respectively:
\begin{align}
\mathbf{h}_{behavior} = \text{Concat}(h_e, h_{\phi(e)}, h_{\psi(e)}).
\end{align}
Sentient clusters system behaviors to provide high-level behavioral representations and performs alert graph correlation. Behaviors are assigned to cluster $C_k$ using:
\begin{align}
C_k = \{e_i | \arg\min_k ||\mathbf{h}_{behavior}^{(i)} - \boldsymbol{\mu}_k||^2\},
\end{align}
where $\boldsymbol{\mu}_k$ represents the centroid of cluster $k$. This enables merging alerts with similar behaviors, thereby alleviating the investigative burden on security analysts.

\section{Evaluation}
\subsection{Datasets and Settings}
\subsubsection{Datasets.}
To comprehensively evaluate our approach, we selected three representative datasets for evaluation:
\textbf{Streamspot}~\cite{streamspot}: This dataset contains five benign scenarios and one malicious scenario, with 100 provenance graphs generated for each scenario. \textbf{Unicorn Wget Dataset}~\cite{Unicorn}: This dataset consists of simulated attacks designed by UNICORN and includes 150 batches of logs collected by CamFlow, with 125 batches being benign and 25 batches malicious. \textbf{DARPA-E3 Dataset}~\cite{darpa_e3}: This dataset was collected from an enterprise network during a defense exercise, where the Red Team conducted Advanced Persistent Threat attacks to exploit vulnerabilities and steal sensitive data. Since no official labels are available, we adopt the ground truth labels used by ThreaTrace and Flash for evaluation.

\subsubsection{Baselines Methods.}
For comprehensive evaluation, Sentient is compared with six state-of-the-art graph-level and node-level methods.
Specialized expert rules or threat reports for APT detection schemes are difficult to compare, so they are not taken into consideration.

\noindent \textbf{1) Graph-level detection}

\noindent\textbf{Streamspot~\cite{manzoor2016streamspot}}: Extracts local graph features through breadth-first search and clusters snapshots to detect anomalies.
\noindent\textbf{Unicorn~\cite{han2020unicorn}}: Analyzes system-wide data provenance graphs in real-time, employing graph sketching techniques and evolutionary models to detect abnormal behavioral patterns in APT attacks.

\noindent \textbf{2) Node-level detection}

\noindent\textbf{Log2vec~\cite{log2vec}}: Constructs heterogeneous graphs from audit logs and applies graph embedding techniques to detect abnormal activities.
\noindent\textbf{Threatrace~\cite{wang2022threatrace}}: Utilizes GraphSAGE to aggregate neighbor node information for learning structural patterns, identifying anomalies based on deviations from learned behaviors.
\noindent\textbf{Flash~\cite{rehman2024flash}}: Integrates Word2Vec semantic encoding with GraphSAGE structural encoding to generate node embeddings for anomaly detection.
\noindent\textbf{Slot~\cite{qiao2025slot}}: Employs graph reinforcement learning to guide provenance graph mining with adaptive neighbor selection for anomaly detection.

\subsubsection{Implementation.} We implemented Sentient in Python 3.10 using PyTorch as the core development framework, with the torch-based DGL framework for graph learning and the Gensim library for Word2Vec. The implementation comprises approximately 3,000 lines of code. All experiments were conducted on a system running Ubuntu 22.04, equipped with an Intel(R) Core(TM) i5-12490F CPU, an NVIDIA GTX 4060Ti GPU, and 64GB of RAM. The final results shown are an average of multiple experiments.

\subsection{Overall Detection Efficacy Comparsion}
The Sentient is trained using a portion of the benign data from the dataset, while the remaining benign data mixed with malicious data is used for testing.
\begin{table}[h]
    \setlength{\tabcolsep}{1mm}
    \centering
    \begin{threeparttable}
    \begin{tabular}{@{}clcccc@{}}
    \toprule
    Datasets & Systems & Precision & Recall & F-score & FPR \\ \midrule
    \multirow{4}{*}{Streamspot} & Streamspot & 73\% & 91\% & 81\% & 6.6\% \\
     & Unicorn & 95\% & 97\% & 96\% & 1.1\% \\
     & Threatrace & \underline{\textit{98\%}} & \underline{\textit{99\%}} & \underline{\textit{98\%}} & \underline{\textit{0.4\%}} \\
     & \textbf{Sentient} & \textbf{99\%} & \textbf{99\%} & \textbf{99\%} & \textbf{0.2\%} \\ \midrule
    \multirow{3}{*}{\makecell{Unicorn \\ Wget}} & Unicorn & 86\% & 95\% & 90\% & 15.5\% \\
     & Threatrace & \underline{\textit{93\%}} & \underline{\textit{98\%}} & \underline{\textit{95\%}} & \underline{\textit{7.4\%}} \\
     & \textbf{Sentient} & \textbf{96\%} & \textbf{99\%} & \textbf{97\%} & \textbf{4.1\%} \\ \midrule
    \multirow{5}{*}{\makecell{DARPA \\ E3 Cadets}} & Log2vec & 49\% & 85\% & 62\% & 3.3\% \\
     & Threatrace & 84\% & 99\% & 91\% & 0.7\% \\
     & Flash & 92\% & \underline{\textit{99\%}} & \underline{\textit{95\%}} & 0.3\% \\
     & Slot & \underline{\textit{94\%}} & 96\% & 95\% & \underline{\textit{0.2\%}} \\
     & \textbf{Sentient} & \textbf{96\%} & \textbf{99\%} & \textbf{97\%} & \textbf{0.2\%} \\ \midrule
    \multirow{5}{*}{\makecell{DARPA \\ E3 Theia}} & Log2vec & 62\% & 66\% & 64\% & 3.2\% \\
     & Threatrace & 79\% & 99\% & 88\% & 2.1\% \\
     & Flash & 91\% & \underline{\textit{99\%}} & 95\% & 0.8\% \\
     & Slot & \underline{\textit{92\%}} & 98\% & \underline{\textit{95\%}} & \underline{\textit{0.7\%}} \\
     & \textbf{Sentient} & \textbf{95\%} & \textbf{99\%} & \textbf{97\%} & \textbf{0.4\%} \\ \midrule
    \multirow{5}{*}{\makecell{DARPA \\ E3 Trace}} & Log2vec & 54\% & 78\% & 64\% & 3.9\% \\
     & Threatrace & 72\% & 99\% & 83\% & 2.3\% \\
     & Flash & 93\% & \underline{\textit{99\%}} & 96\% & 0.4\% \\
     & Slot & \underline{\textit{94\%}} & 98\% & \underline{\textit{96\%}} & \underline{\textit{0.4\%}} \\
     & \textbf{Sentient} & \textbf{97\%} & \textbf{99\%} & \textbf{98\%} & \textbf{0.2\%} \\ \bottomrule
    \end{tabular}
    \begin{tablenotes}[flushleft]
    \item \textbf{Bold} indicates the best performance. 
    \item \underline{\textit{Italic underline}} indicates the second-best performance.
    \end{tablenotes}
    \caption{Performance Comparison of Sentient with State-of-the-Art Methods at Graph-level and Node-level.}
    \label{performance_comparison}
    \end{threeparttable}
\end{table}

As shown in Table~\ref{performance_comparison}, we evaluated Sentient's detection efficacy at the graph-level using the Streamspot and Unicorn Wget datasets, comparing it with Streamspot, Unicorn, and Threatrace. In the Streamspot dataset, which involves relatively simple attack scenarios, Sentient easily detected anomalies and achieved near-perfect performance. On the Unicorn Wget dataset, complex scenarios compromised the detection efficacy of other methods, while Sentient maintained robust performance.

We evaluated Sentient using the DARPA E3 dataset against other node-level methods (Log2vec, Threatrace, Flash, and Slot). Sentient achieved superior results due to its capability to perceive indirect dependence in complex scenarios, reduce environmental noise, and capture logical associative relationships. In contrast, GNN-based methods (such as Threatrace, Flash, and Slot) that aggregate neighbor information fail to establish logical relationship correlations for behaviors in complex scenarios.

\subsection{Performance Overhead}
\begin{figure}[h]
    \subfigure[Time Overhead]{
        \includegraphics[width=0.22\textwidth]{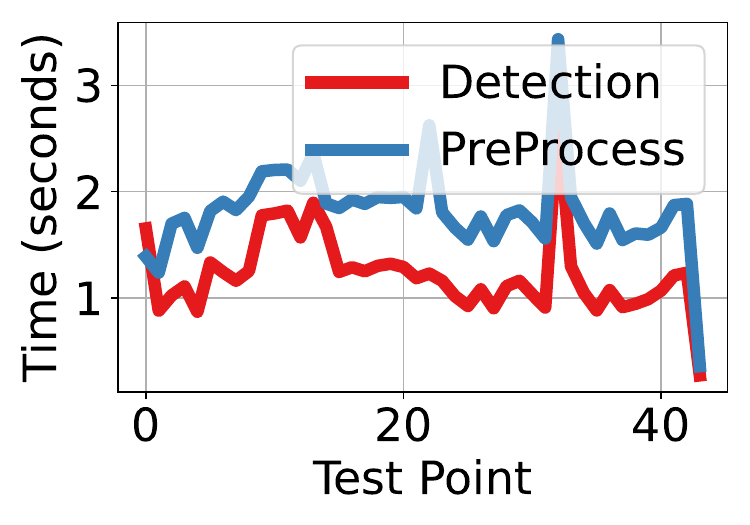}
        \label{fig:overhead1}
    }
    \subfigure[Memory Usage]{
        \includegraphics[width=0.22\textwidth]{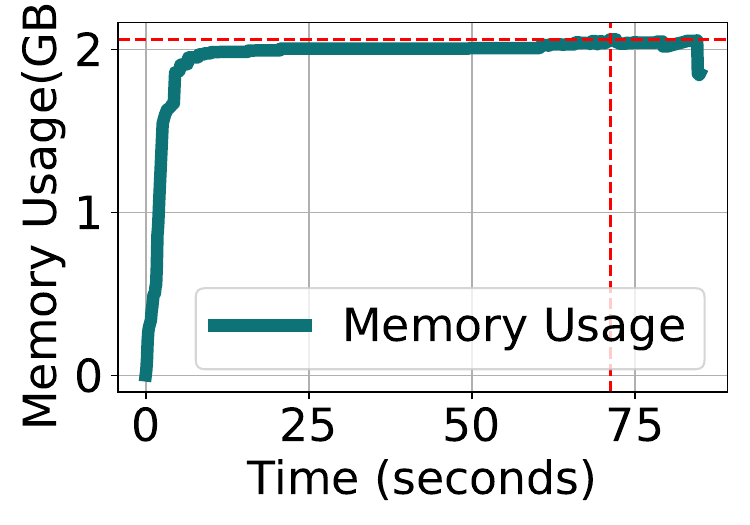}
        \label{fig:overhead2}
    }
    \caption{Performance overhead of Sentient on the Cadets.}
    \label{fig:overhead}
\end{figure}
We evaluated Sentient's computational overhead using the DARPA E3 Cadet dataset, with results averaged over multiple experimental runs. Figure \ref{fig:overhead1} demonstrates the time costs for preprocessing (raw log to provenance graph conversion) and detection phases, averaging 1.82 seconds and 1.23 seconds respectively. Figure \ref{fig:overhead2} shows Sentient's memory consumption during detection, with peak usage of 2.01GB. Given that the E3-Cadet dataset generates approximately 2.6GB of audit logs daily over a two-week collection period, our experiments on two days of log data show that processing one day's logs (including preprocessing) requires only 63.6 seconds on average. This indicates that Sentient can meet the daily detection requirements of small organizations and enterprises with acceptable performance overhead.

\subsection{Ablation Study}
\begin{figure}[h]
    \subfigure[Detection Performance]{
        \includegraphics[width=0.22\textwidth]{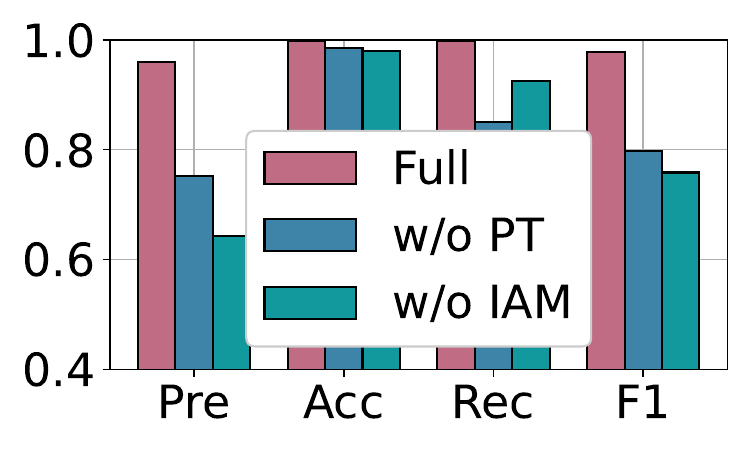}
        \label{fig:Ablation1}
    }
    \subfigure[Overhead]{
        \includegraphics[width=0.22\textwidth]{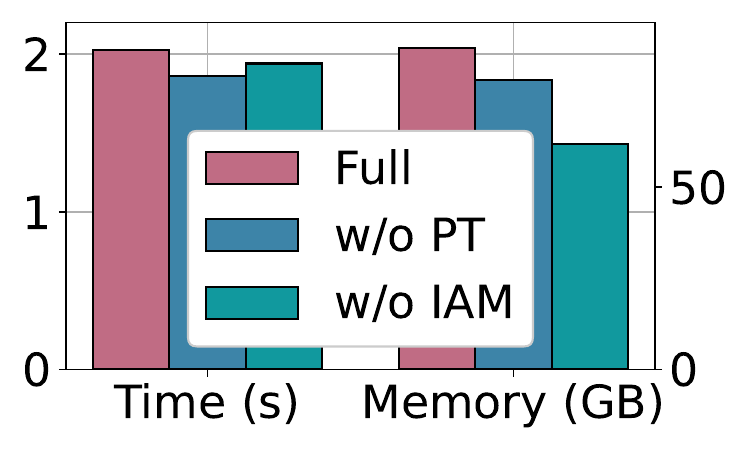}
        \label{fig:Ablation2}
    }
    \caption{Ablation Study of PT and IAM Components on CADETS Dataset: Impact on Performance and Overhead.}
    \label{fig:Ablation}
\end{figure}

We conducted an ablation study of Sentient using the Cadets dataset to evaluate the importance of different components, as shown in Figure~\ref{fig:Ablation}.

\noindent \textbf{Impact of Pre-training (PT)}: Removing Sentient's PT component, which combines node attribute and position encoding as node embedding, caused precision to drop by 20.75\%. This indicates that PT capturing complex interaction information (e.g., indirect dependencies) is crucial for node representation.

\noindent \textbf{Impact of Intent Analysis Module (IAM)}: Removing the Bi-Mamba2 model from IAM, which directly feeds sequence representation to the MLP for action reconstruction, led to a 31.59\% decrease in precision. This demonstrates that learning behavioral correlations is crucial for accurate threat detection.

\subsection{Hyperparameter Impact on Performance}
\begin{figure}[h]
    \subfigure[Embedding Dimension]{
        \includegraphics[width=0.22\textwidth]{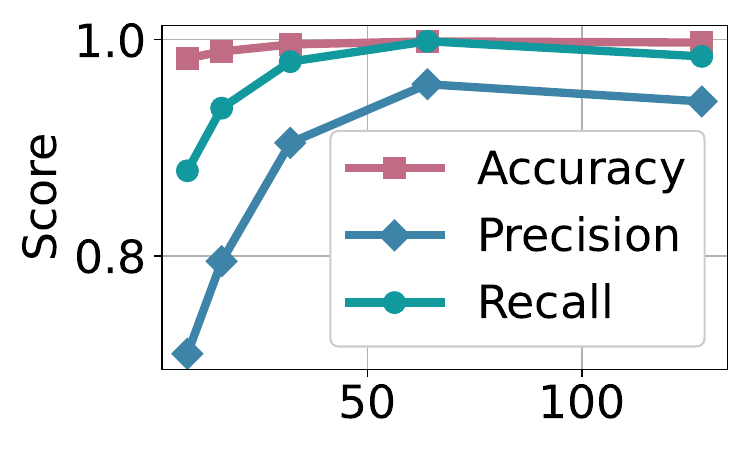}
        \label{fig:parameter1}
    }
    \subfigure[Sequence Length]{
        \includegraphics[width=0.22\textwidth]{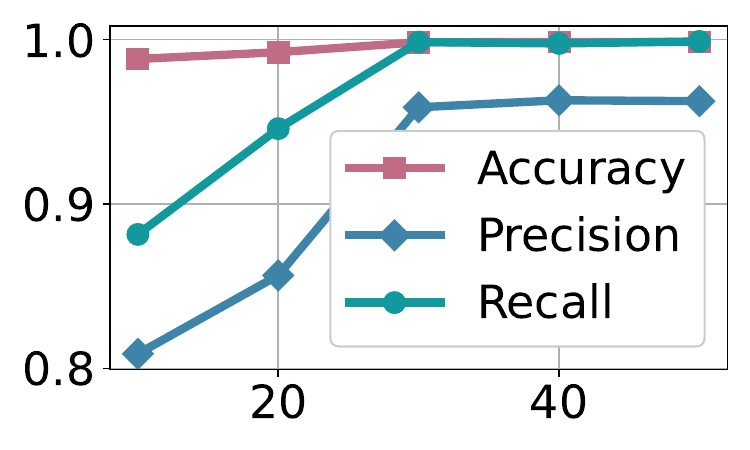}
        \label{fig:parameter2}
    }
    \caption{Hyperparameter Impact on Performance.}
    \label{fig:parameter}
\end{figure}

In previous sections, we evaluated Sentient with a set of fixed hyperparameters. Here, we conduct experiments on the Cadets dataset, varying key parameters to assess their impact on Sentient's performance, as shown in Figure~\ref{fig:parameter}.

\noindent \textbf{Node Embedding Dimension}: Sentient maps structural and attribute information of nodes into embeddings during the graph learning phase. The embedding dimension affects how well the embeddings represent node information. We explored different embedding dimensions, as shown in Figure~\ref{fig:parameter1}. Lower dimensions reduce expressive power, negatively impacting detection, while higher dimensions lead to sparse features. We found that 64 dimensions yielded optimal performance.

\noindent \textbf{Sequence Length}: Sequence length determines the receptive field of events. As shown in Figure~\ref{fig:parameter2}, longer sequences offer a broader receptive field but also increase performance overhead. Performance gains plateau beyond a length of 30. Considering both benefits and costs, we found 30 to be the optimal length. In practice, behavioral sequence lengths rarely exceed 30.

\subsection{Robustness Against Mimicry Attacks}
\begin{table}[h]
\setlength{\tabcolsep}{3.3mm}
\centering
\begin{tabular}{@{}ccccc@{}}
\toprule
\makecell{\textbf{Adversarial} \\ \textbf{Events}} & \textbf{Precision} & \textbf{Recall} & \textbf{F-score} \\ \midrule
None  & \makecell{95.88\% } & \makecell{99.85\% } & \makecell{97.83\% } \\ 
\midrule
1000 Events & \makecell{95.39\% \\ (↓0.49\%)} & \makecell{99.00\% \\ (↓0.85\%)} & \makecell{97.16\% \\ (↓0.67\%)} \\
2000 Events & \makecell{94.96\% \\ (↓0.92\%)} & \makecell{98.93\% \\ (↓0.92\%)} & \makecell{96.90\% \\ (↓0.93\%)} \\
3000 Events & \makecell{95.52\% \\ (↓0.35\%)} & \makecell{99.17\% \\ (↓0.68\%)} & \makecell{97.31\% \\ (↓0.52\%)} \\
\bottomrule
\end{tabular}
\caption{Adversarial Mimicry Attack Against Our System with Varying Numbers of Added Events. Values in parentheses indicate relative changes compared to our method without adversarial data.}
\label{tab:Adversarial}
\end{table}
We evaluated Sentient's resilience against adversarial attacks using the mimicry attack method from \cite{goyal2023sometimes}, which adds benign structures to malicious nodes to conceal malicious behaviors. Experiments on the Cadet dataset with varying numbers of added benign structures demonstrate that mimicry attacks have minimal negative impact on Sentient's performance, as shown in Table~\ref{tab:Adversarial}. This robustness stems from Sentient's ability to leverage global contextual information through pre-trained models and learn logical behavioral correlations via neighbor denoising.  While mimicry attacks may deceive most methods by improving the overall impression of malicious nodes, they cannot conceal the inherent logical correlations of attack behaviors.

\subsection{Alert Validation}
\begin{figure}[h]
    \centering
    \includegraphics[width=\linewidth]{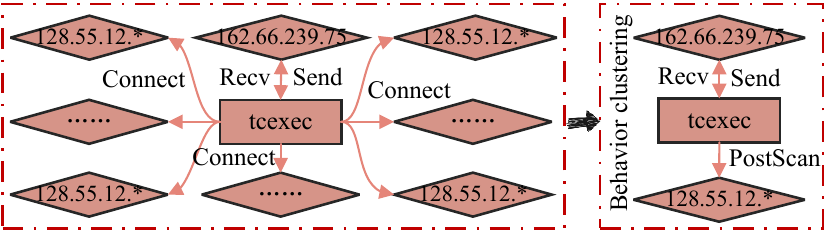}
    \caption{An example of attack activities reconstructed.}
    \label{fig:Alert}
\end{figure}

Sentient aims to provide security analysts with effective and concise alerts to accelerate threat response.  As demonstrated in previous evaluation, Sentient's superior performance in reducing false positives and false negatives significantly decreases analysts' workload.  Additionally, during attack investigation, Sentient clusters behaviors with similar intents to further simplify attack chains.  For instance, the port scanning trace shown in Figure~\ref{fig:Alert} involves 67,374 nodes and 134,545 behaviors.  After Sentient's behavioral intent clustering, analysts receive the simplified attack graph illustrated in the right of Figure~\ref{fig:Alert}.

\section{Conclusion}
Advanced Persistent Threats pose significant cybersecurity challenges due to their stealthy and sophisticated nature. Existing detection methods fail to capture indirect dependencies and logical correlations between behaviors, resulting in high false positive rates and overlooking stealthy attacks. We propose Sentient, an APT detection method that combines graph transformers for capturing indirect dependencies with bidirectional Mamba2 for learning behavioral correlations. Sentient requires only benign data for training and detects anomalies that deviate from normal behavioral patterns. Evaluation on three widely-used datasets demonstrates superior performance, achieving a 44\% average reduction in FPR compared to state-of-the-art methods.

\appendix
\section*{Acknowledgments}
This research is supported by National Key Research and Development Program of China (No.2023YFC2206402), the Strategic Priority Research Program of the Chinese Academy of Sciences (No.XDA0460100), and Youth Innovation Promotion Association CAS (No.2021156). This work is also supported by the Program of Key Laboratory of Network Assessment Technology, the Chinese Academy of Sciences, Program of Beijing Key Laboratory of Network Security and Protection Technology.

\bibliography{aaai2026}

\end{document}